\documentclass[12pt]{article}
\usepackage{epsfig}

\usepackage{xcolor}

\usepackage{a4}
\usepackage{amsmath}

\usepackage{amssymb}
\usepackage{graphicx}
\usepackage{amsmath}
\usepackage{amsfonts} 

\newcommand{\la}{\lambda}

\newcommand{\f}{\phi}

\newcommand{\si}{\sigma}

\newcommand{\ee}{\end{equation}}
\newcommand{\eea}{\end{eqnarray}}
\newcommand{\be}{\begin{equation}}
\newcommand{\bea}{\begin{eqnarray}}

\newcommand{\pa}{\partial}
\newcommand{\Om}{\Omega}
\newcommand{\vep}{\varepsilon}
\newcommand{\vr}{\varrho}

\newcommand{\om}{\omega}
\def\theequation{\arabic{equation}}

\newcommand{\re}[1]{(\ref{#1})}
\newcommand{\R}{{\rm I \hspace{-0.52ex} R}}

\newcommand{\eins}{1\hspace{-0.56ex}{\rm I}}

\def\theequation{\thesection.\arabic{equation}}
\numberwithin{equation}{section}
\date{}

\begin{document}

\title{\bf Gauged Skyrme analogue of Chern-Pontryagin}

\author{ {\large D. H. Tchrakian}$^{\dagger *}$
\\  
\\ 
$^{\dagger}${\small 
School of Theoretical Physics, Dublin Institute for Advanced Studies,}
\\
{\small  Burlington Road, Dublin 4, Ireland}
\\   
$^{*}${\small Department of Computer Science, National University of Ireland Maynooth, Maynooth, Ireland }
}

\maketitle

\date{}

 \begin{abstract}
An analogue of the Chern-Pontryagin density for $SO(D)$ gauged $O(D+1)$ Skyrme systems,
referred to as Skyrme--Chern-Pontryagin (SCP) densities is known for dimensions $D=2,3,4$. Since these
are defined only through a prescription, it is necessary to extend the realisation to higher $D$, which
is carried out here for $D=5$. 
The construction of SCP densities in $D=2,3,4,5$ is presented here in a unified pattern with
the aim of pointing out to the possible extrapolation to all dimensions.
\end{abstract}

\bigskip
\bigskip
\bigskip
\bigskip
\bigskip
\bigskip
\bigskip

\bigskip
\bigskip
\bigskip
\bigskip
\bigskip
\bigskip
\bigskip

\noindent
{\it email of author:} tigran@stp.dias.ie



\newpage

\section{Introduction}
Chern-Pontryagin (CP) densities are pivotal in the construction of instantons~\cite{Belavin:1975fg} on $\R^4$,
and their dimensional descendants play that role for the magnetic monopoles~\cite{tHooft:1974kcl} on $\R^3$
and the Abelian Higgs vortices~\cite{Nielsen73} on $\R^2$. CP densities play the same role also in higher
dimensions. In even dimensions, they stabilise Yang-Mills instanton actions and in particular on $\R^{4p}$
they support (anti-)self-dual instantons~\cite{Tchrakian:1984gq}. Their dimensional descendants are
employed in constructing magnetic monopoles~\cite{Tchrakian:2010ar} in all odd and even dimensions.

Concerning gauged Skyrmions the situation is entirely different since Skyrme scalars are not dimensional
descendants of Yang-Mills fields in higher (even) dimensions, and hence CP densities cannot be employed in
the same way as for gauged Higgs fields\footnote{This said, it should be noted that group valued Skyrme
fields $U(x)$ are the holonomy of Yang-Mills fields. This is best known in $3$ 
dimensions~\cite{Atiyah:1989dq} where the $O(4)$
Skyrme scalar $\f^a\ ;\ a=1,\dots,4$ encodes the $SU(2)$ group element $U=\f^a\tau^a\ ,\ U^{-1}=\f^a\tilde\tau^a$
; $\tau^a=(i\vec\si,\eins)\ , \tilde\tau^a=(-i\vec\si,\eins)$.}

Another very important difference between gauged Higgs fields and gauged Skyrmions is,
that choice of the gauge group of the former is restricted by the condition of asymptotic constancy of the
covariant derivative of the Higgs field, namely to $SO(D)$ for $\R^D$, while a $O(D+1)$ Skyrmion can be gauged
with any group $SO(N)$ as long as $N\le D$.

The approach taken here is to devise an analogue of the CP
density that accommodates the Skyrme scalar in the given dimension.

The guiding principle for devising a density that can present a lower bound for the action/energy for a
gauged Skyrme system is to find an analogue of the CP density, which like the latter is
both {\it gauge invariant} and {\it total divergence}. This then
can be exploited to establish Bogomol'nyi like action/energy lower bounds.

Formally, this density, $\vr^{(D)}$ in $D$ dimensions is expressed as
\bea
\vr^{(D)}&=&\vr_G^{(D)}+W[F,D\f]\label{vr1}\\
&=&\vr_0^{(D)}+\pa_i\Om_i[A,\f]\,,\label{vr2}
\eea
where
\bea
\vr_0^{(D)}&=&\vep_{i_1i_2\dots i_D}\vep^{A_1A_2\dots A_DA_{D+1}}\pa_{i_1}\f^{A_1}\pa_{i_2}\f^{A_2}\dots\pa_{i_D}\f^{A_D}\f^{A_{D+1}}
\label{vr0}\\
\vr_G^{(D)}&=&\vep_{i_1i_2\dots i_D}\vep^{A_1A_2\dots A_DA_{D+1}}D_{i_1}\f^{A_1}D_{i_2}\f^{A_2}\dots D_{i_D}\f^{A_D}\f^{A_{D+1}}
\label{vrG}
\eea
and where $\vr_0^{(D)}$ defined by \re{vr0} is the {\it winding number density} of the $O(D+1)$ Skyrme scalar $\f^A\ ,\ A=1,2,\dots,D+1$ in $D$ dimensions,
subject to the constraint $|\f^A|^2=1$, and $\vr_G^{(D)}$ is defined by replacing all the partial derivatives in $\vr_0^{(D)}$ by
covariant derivatives. Thus, $\vr_0$ is effectively a total-divergence but is not gauge-invariant, while $\vr_G^{(D)}$ is by
construction gauge-invariant but is not total-divergence.

In \re{vr0} and \re{vrG}, coordinates are labelled by lower indices, $e.g.,$ $x_i$, and the indices
labelling the $O(D+1)$ scalars by upper indices,  $e.g.,$ $\f^A$. Upper/lower here do not imply
contravariant/covariant. This notation is used throughout.

For $\vr^{(D)}$ to present an energy lower bound it must be both {\it gauge invariant}
and {\it total divergence}, which is achieved by constructing the gauge-invariant quantity $W[F,D\f]$
and the total-divergence $\pa_i\Om_i[A,\f]$. In this sense, $\vr$ is
an analogue of the Chern-Pontryagin (CP), which is the density used in establishing action/energy lower bounds for
instantons/monopoles (and vortices). In the case of instantons, this is the CP density itself, while in the case of
monopoles (and vortices) it is the appropriate dimensional descendant of a CP density.

Unlike the CP density however, $\vr^{(D)}$ in $D$ dimensions, prescribed by \re{vr1}-\re{vr2},
is defined for gauge group $SO(D)$, or contractions to a subgroup of $SO(D)$. To make the prescription
\re{vr1}-\re{vr2} concrete, it is necessary to define the gauging prescription and fix the notation.

The $O(D+1)$ Skyrme scalar is $\f^A\ ,\ A=1,2,\dots,D+1$ is subject to
\be
|\f^A|^2=|\f^a|^2+(\f^{D+1})^2=1\ ,\quad a=1,2,\dots,D\,.\label{constr}
\ee

The $SO(D)$ gauging prescription is
\bea
D_i\f^a&=&\pa_i\f^a+A_i\f^a\ ,\quad A_i\f^a=(A_i\f)^b=A_i^{ab}\f^b\label{coval22}\\
D_i\f^{D+1}&=&\pa_i\f^{D+1}\,,\label{covaA22}
\eea
where
\[
A_i^{ab}=-A_i^{ba}\ \ i=1,2,\dots,D\ ;a=1,2,\dots,D 
\]
is the $SO(D)$ gauge connection, whose curvature is
\bea
F_{ij}^{ab}&=&\pa_{[i}A_{j]}^{ab}+(A_{[i}A_{j]})^{ab}\label{curv1}\\
&\equiv&\pa_iA_j^{ab}+A_i^{ac}A_j^{cb}-(i,j)\nonumber\,.
\eea

The prescription for constructing the \re{vr1}-\re{vr2} is to calculate the difference
\be
\label{rGminusr0}
\vr_G^{(D)}-\vr_0^{(D)}=\pa_i\Om_i[A,\f]-W[F,D\f]
\ee
such that $W[F,D\f]$ is gauge invariant.

The relation \re{rGminusr0} is evaluated explicitly by extracting partial derivatives until a gauge invariant quantity is isolated.
In increasing dimensions, there is some arbitrariness in this splitting. The
calculations are carried out directly, using the Leibniz rule, tensor and Bianchi identities.

As stated above, the density \re{vr1}-\re{vr2},
\bea
\vr^{(D)}&=&\vr_G^{(D)}+W[F,D\f]\nonumber\\
&=&\vr_0^{(D)}+\pa_i\Om_i[A,\f]\,,\nonumber
\eea
is an analogue of the Chern-Pontryagin (CP) density and can therefore be similarly exploited.
It might not be unreasonable to refer to $\vr^{(D)}$ as a Skyrme--Chern-Pontryagin (SCP) density.

The definition \re{vr1}  of the SCP density is suited for devising Bogomol'nyi type energy lower bounds,
has been exploited in constructing gauged Skyrme solitons.
It was first proposed in Ref.~\cite{Schroers:1995he} for Maxwell dynamics in
the case $D=2$ which served as the template for extending
to higher dimensions. Also in $D=2$ this method was adapted to Chern-Simons dynamics in \cite{Arthur:1996uu}.
It was subsequently extended to $\R^3$ in \cite{Arthur:1996np,Brihaye:2001qt,Brihaye:2000zsq} and $\R^4$ in \cite{Brihaye:2000zsq}.

More recently gauged Skyrmions in $D=3$ were also constructed. This was done for Maxwell gauging in
\cite{Livramento:2023tmm}, and for $SU(2)$ in~\cite{Cork:2018sem,Cork:2021ylu,Cork:2023pft}, where the main aim was to
construct Skyrmions whose energy lower bounds are closer~\footnote{Introducing a new field in a model results in lower energy,
in this case the gauge field being the new field.
this was observed also in \cite{Brihaye:2000zsq} for $SO(3)$ and in \cite{Piette:1997ny} for $SO(2)$.}
to the (topological) ``baryon number''. The energy densities in these models feature further gauge invariant terms in addition to
those appearing in \cite{Brihaye:2000zsq,Piette:1997ny}. In \cite{Cork:2018sem} the relation to calorons and monopoles is
exploited to give approximate $SU(2)$ gauged Skyrmions, while in~\cite{Cork:2021ylu} the symmetry was further broken down
to $U(1)$ to approach the lower bound further. In~\cite{Cork:2023pft} equivariant cohomology is used to formulate the
topological charge of the $SU(2)$ gauged Skyrmion.

The definition \re{vr2} for the SCP density is useful for
calculating the effective ``baryon number'' which in the presence of the Chern-Simons term can cause this to
depart from the {\it integer valued} topological charge. This was observed~\footnote{This feature is restricted to
odd dimensional spacetimes where it is possible to express the gauged Skyrmion in terms of a radial variable, which results from
subjecting the multi-azimuthal fields to further symmetry constraining it to one-dimension. This is clearly not the case in
$3+1$ dimensions as seen in \cite{Brihaye:2001qt}, where this question remains open.}
in $2+1$ and $4+1$ dimensional
Abelian gauged $O(3)$ and $O(5)$ models, in
\cite{Navarro-Lerida:2018giv} and \cite{Navarro-Lerida:2020hph}, respectively.

Not less importantly, since \re{vr2} can be expressed explicitly as a total divergence, it can be exploited to define the
Skyrme--Chern-Simons (SCS) density proposed in Refs.~\cite{Tchrakian:2015pka,Tchrakian:2021xzy}. This is a Chern-Simon like density
whose definition will be given below in Section {\bf 3}.

However, unlike the CP density which can be defined for any gauge group in any even dimension, the gauge group of 
the SCP density in $D$ dimensions, prescribed by \re{vr1}-\re{vr2}, is restricted to be at most $SO(D)$, or some subgroup of it.
But unlike the the CP density which can be defined in even dimensions only, the definition of the SCP density in {\bf not} 
restricted to even dimensions only and is valid in odd dimensions too.

We are now confronted with the essential question. While the definition of the CP density for any gauge group and any
(even) dimension is unambiguously stated, the corresponding status of the SCP density depends on explicit construction in any given
dimension, and as such \re{vr1}-\re{vr2} is a prescription. To date, this prescription was implemented only for $D=2,\ 3$ and $4$,
and was exploited in Refs.~\cite{Arthur:1996uu}, \cite{Arthur:1996np} and \cite{Brihaye:2000zsq} respectively.

To answer this question at least partially, it is necessary to explore the prescription \re{vr1}-\re{vr2} in dimensions
$D\ge 5$, and in the present note this carried out for dimension $D=5$. This is the most important aim of the present work. In
Section {\bf 2} below we present the construction of $\vr^{(D)}$ for $D=2,\ 3,\ 4$ and $5$, in successive Subsections.
The aim is to give an as transparent as possible description of these SCP densities in a unified manner. In Section {\bf 3},
a summary and outlook is given, and the further details of the content of Section {\bf 2} are relegated to the Appendix.

\section{Calculation}
The quantity $(\vr_G^{(D)}-\vr_0^{(D)})$ is calculated and cast in the prescribed form of \re{rGminusr0},
for $D=2,3,4,$ and $5$ in each Subsection below, respectively.

\subsection{$D=2$}
Using basic tensor identities, $(\vr_G-\vr_0)$ given by \re{vrG} minus \re{vr0}, is unpacked to give
\be
\frac12(\vr_G-\vr_0)=\frac12\,\vep_{ij}\,\vep^{fg}\,\pa_j\f^3\ A_i^{fg}
\equiv\vep_{ij}\,\pa_j\f^3\ A_i\,,\label{rGr02}
\ee
which by a single application of the Leibniz rule splits in a gauge-invariant and a total divergence
\be
\label{rGr02x}
(\vr_G-\vr_0)=\vep_{ij}\left[2\,\pa_j(\f^3\ A_i)-\f^3\,F_{ij}\right]\,,
\ee
from which the quantities $W$ and $\pa_i\Om_i$ in \re{vr1} and \re{vr2} can be read off
\bea
W&=&\vep_{ij}\,\f^3\,F_{ij}\label{W2}\\
\Om_i&=&-2\vep_{ij}\,\f^3\,A_j\label{Om2}
\eea

Of course, in all even $D$ the $(D/2)$-th CP density can be added to both \re{vr1} and to \re{vr2}.
preserving the CP like properties of $\vr^{(2)}$.
For $D=2$, this is the $1$-st CP density, with  $\vep_{ij}F_{ij}$ being subtracted from $W$ in \re{W2} and
$2\vep_{ij}A_j$ added to $\Om_i$ in \re{Om2}.

\subsection{$D=3$}
Using basic tensor identities, $(\vr_G-\vr_0)$ given by \re{vrG} minus \re{vr0}, is unpacked to give
\be
\frac13(\vr_G-\vr_0)=\vep_{ijk}\,\vep^{fga}\ \pa_k\f^4\ A_j^{fg}
\left(\pa_i\f^a+\frac12 A_i\f^a\right)\,.\label{rGr03}
\ee

Applying the Leibniz rule, \re{rGr03} can be written as
\be
\label{result3}
\frac13(\vr_G-\vr_0)=\pa_k\f^4\left(\pa_i\Om_{ik}-\frac12\,\vep_{ijk}\vep^{fga}\,\f^a\,F_{ij}^{fg}\right)
\ee
where
\be
\label{Omik}
\Om_{ik}=\vep_{ijk}\vep^{fga}\,(A_j^{fg}\,\f^a)\,.
\ee

The result \re{result3} is of the required form \re{rGminusr0}
 as the first term $\pa_k\f^4\pa_i\Om_{ik}$ is total divergence by virtue of the antisymmetry of $\Om_{ik}$,
 and, the second term is gauge invariant as it stands.

From the viewpoint of the application of Bogomol'nyi inequalities however, it is expedient to modify this
result by extracting the partial derivative from $\pa_k\f^4$ in the gauge invariant term
in \re{result3}. Using the Bianchi identity, this
results in the identity
\be
\label{ident3}
\vep_{ijk}\,\vep^{fga}\,\pa_k\f^4\,\f^a\,F_{ij}^{fg}=\vep_{ijk}\,\vep^{fga}\left[\pa_k\left(\f^4\,\f^a\,F_{ij}^{fg}\right)-
\f^4\,D_k\f^a\,F_{ij}^{fg}\right]
\ee
yielding the result in the desired form
\bea
W&=&\frac32\,\vep_{ijk}\vep^{fga}\f^4\,D_k\f^a\,F_{ij}^{fg}\label{W3}\\
\Om_k&=&3\,\f^4\left(\pa_i\Om_{ik}-\frac12\,\vep_{ijk}\vep^{fga}\,\f^a\,F_{ij}^{fg}\right)\,.\label{Omk3}
\eea

Exploiting identities like \re{ident3} resulting from the extraction of a partial derivative are a persistent
feature in achieving the desired form of $W$ and $\Om_i$, in all dimensions $D\ge 3$.
\bea
\Om_k&=&3\,\f^4\left(\pa_i\Om_{ik}-\frac12\,\vep_{ijk}\vep^{fga}\,\f^a\,F_{ij}^{fg}\right)
\Om_{ik}=\vep_{ijk}\vep^{fga}\,(A_j^{fg}\,\f^a)\nonumber
\eea

\subsection{$D=4$}
Using basic tensor identities, $(\vr_G-\vr_0)$ given by \re{vrG} minus \re{vr0}, is unpacked to give
\be
\frac14(\vr_G-\vr_0)=\vep_{ijkl}\vep^{fgab}\pa_l\f^5\,A_k^{fg}\left(\pa_i\f^a\pa_j\f^b+\pa_i\f^aA_j\f^b+\frac13\,A_i\f^aA_j\f^b\right)\,.\label{rGr04}
\ee

After multiple applications of the Leibniz rule and tensor identities, this can be put in the form
\be
\frac14(\vr_G-\vr_0)=\pa_l\f^5\pa_i\Om_{il}+\frac12\,\vep_{ijkl}\vep^{fgab}\pa_l\f^5\left\{
F_{ij}^{fg}\,\f^aD_k\f^b+\frac12|\vec\f|^2\Xi_{ijk}^{fgab}
\right\}\label{vrv04}
\ee
where we have used the notation $|\vec\f|^2=|\f^a|^2$, and where
\bea
\Om_{il}&=&\vep_{ijkl}\vep^{fgab}A_k^{fg}\,\f^a
\left(\pa_j\f^b-\frac12A_j\f^b\right)\,,\label{tildeOm4}
\eea
and
\be
\label{Xi4}
\Xi_{ijk}^{fgab}=A_k^{ab}\left[\pa_iA_j^{fg}+\frac23(A_iA_j)^{fg}\right]\,.
\ee

Note that the $l=4$ component of $\vep_{ijkl}\vep^{fgab}\,\Xi_{ijk}^{fgab}$, namely
\[
\vep_{\mu\nu\la 4}\vep^{fgab}\,\Xi_{\mu\nu\la}^{fgab}\ ,\quad \mu=1,2,3\ ,
\]
is just the (Euler--)Chern-Simons density in three dimensions.

The expressions \re{vrv04} and \re{tildeOm4} are the analogues of \re{result3} and \re{Omik} above in the $D=3$
case. Again, the first term in \re{vrv04}, $\pa_l\f^5\pa_i\Om_{il}$, is a total divergence by virtue of the
antisymmetry of $\Om_{il}$, and the second term is manifestly gauge invariant. But the third term involving
$\Xi_{ijk}^{fgab}$ is neither gauge-invariant nor is it total-divergence. Thus, \re{vrv04} is not of the form
\re{rGminusr0} as it stands. This is a
feature of $(\vr_G-\vr_0)$ for all $D\ge 4$.

To render \re{vrv04} to the form \re{rGminusr0}, namely consisting of a gauge invariant density $W$ and a
total divergence, the partial derivative $\pa_l$ from the gauge-variant
term in \re{vrv04} must be extracted using the Bianchi identities. 
But as above in $D=3$, it is expedient to extract this partial derivative
also from the gauge-invariant term~\footnote{This is convenient from the viewpoint of applying Bogomol'nyi type
inequalities.}.

For the gauge-invariant term this is implemented by
\bea
\vep_{ijkl}\vep^{fgab}\pa_l\f^5\,\{F_{ij}^{fg}\,\f^aD_k\f^b\}&=&\vep_{ijkl}\vep^{fgab}\bigg\{\pa_l
\left[\f^5\,F_{ij}^{fg}\,\f^aD_k\f^b\right]\nonumber\\
&&\qquad\qquad-\f^5\left[F_{ij}^{fg}D_k\f^aD_l\f^b+\frac18|\vec\f|^2F_{ij}^{fg}F_{kl}^{ab}\right]\bigg\}
\label{id1}
\eea
and for the gauge-variant term by
\bea
\vep_{ijkl}\vep^{fgab}|\vec\f|^2\pa_l\f^5\,\Xi_{ijk}^{fgab}
&=&\vep_{ijkl}\vep^{fgab}\bigg\{\pa_l\left[(\f^5-\frac13(\f^5)^3)\,\Xi_{ijk}^{fgab}\right]
+\frac14(\f^5-\frac13(\f^5)^3)F_{ij}^{fg}F_{kl}^{ab}\bigg\}\,.\label{id2}
\eea

Applying the identities \re{id1} and \re{id2}, \re{vrv04} is cast in the required form \re{rGminusr0} with
\bea
W&=&\frac12\,\vep_{ijkl}\vep^{fgab}\,\f^5\bigg[F_{ij}^{fg}D_k\f^aD_l\f^b+
\frac{1}{12}(\f^5)^2\,F_{ij}^{fg}F_{kl}^{ab}\bigg]\label{W4}\\
\Om_l&=&\f^5\bigg(\pa_i\Om_{il}+\frac12\vep_{ijkl}\vep^{fgab}\bigg[F_{ij}^{fg}\f^aD_k\f^b
+\frac12\left(1-\frac13(\f^5)^2\right)\Xi_{ijk}^{fgab}\bigg]\bigg)\,,\label{Oml4}
\eea
which are the $D=4$ analogues of $W$ and $\Om_i$, \re{W3} and \re{Omk3} in $D=3$, it being understood that
the indices $(i,a)$ run over $(1,2,3)$ in \re{W3}-\re{Omk3} while here they run over $(1,2,3,4)$ in 
\re{W4}-\re{Oml4}.

\subsection{D=5}
Using basic tensor identities, $(\vr_G-\vr_0)$ given by \re{vrG} minus \re{vr0}, is unpacked to give
\bea
\frac15(\vr_G-\vr_0)&=&\vep_{ijklm}\,\vep^{fgabc}\ \pa_m\f^6\ A_l^{fg}
\bigg(2\pa_i\f^a\pa_j\f^b\pa_k\f^c+3\pa_i\f^a\pa_j\f^bA_k\f^c\nonumber\\
&&\qquad\qquad\qquad\qquad\qquad\qquad+2\pa_i\f^aA_j\f^bA_k\f^c+\frac12\,A_i\f^aA_j\f^bA_k\f^c\bigg)
\label{rGr05}
\eea

After multiple applications of the Leibniz rule and tensor identities, \re{rGr05} can be put in the form
analogous to its $D=4$ counterpart \re{vrv04}
\bea
\frac15(\vr_G-\vr_0)&=&\pa_m\f^6\pa_i\Om_{im}-\vep_{ijklm}\vep^{fgabc}\pa_m\f^6\left\{
F_{ij}^{fg}\,\f^a\,D_k\f^bD_l\f^c+|\vec\f|^2\,\Xi_{ijkl}^{fgabc}\right\}\label{vrv05}
\eea
where
\bea
\label{tildeOm5}
\Om_{im}
&=&2\vep_{ijklm}\vep^{fgabc}\,A_l^{fg}\,
\f^a(\pa_j\f^b\pa_k\f^c+\pa_j\f^bA_k\f^c+\frac23\,A_j\f^bA_k\f^c)
\eea
and
\bea
\Xi_{ijkl}^{fgabc}&=&A_k^{ab}\left[\left(\pa_l\f^c+\frac23\,A_l\f^c\right)\,\pa_iA_j^{fg}
+\frac23\left(\pa_l\f^c+\frac34\,A_l\f^c\right)(A_iA_j)^{fg}\right]\,,\label{CS5}
\eea
$\Om_{im}$ in \re{tildeOm5} being the $D=5$ analogue of \re{tildeOm4}, and $\Xi_{ijkl}^{fgabc}$ in \re{CS5}
the $D=5$ analogue of $\Xi_{ijk}^{fgab}$ in \re{Xi4},  in $D=4$.

Again, the first term $\pa_m\f^6\pa_i\Om_{im}$ in \re{vrv05} is a total divergence and the second term is
manifestly gauge invariant, while the third term involving $\Xi_{ijkl}^{fgabc}$ is neither gauge invariant nor
total divergence. Thus, to cast \re{vrv05} in the required form consisting of a gauge invariant part $W$ and a
total divergence, the partial derivative $\pa_m$ must be extracted  from this term by using the Bianchi
identities. As in $D=3$ and $4$, it is expedient to extract $\pa_m$ also from the gauge invariant term in
\re{vrv05}.

Extracting the partial derivative $\pa_m$ from the gauge invariant term in \re{vrv05} results in the expression
\bea
&&\vep_{ijklm}\vep^{fgabc}\pa_m\f^6\,F_{ij}^{fg}\,\f^a\,D_k\f^bD_l\f^c=\nonumber\\
&&\qquad\qquad=
\vep_{ijklm}\vep^{fgabc}\bigg\{\pa_m\left[\f^6\,F_{ij}^{fg}\f^aD_k\f^bD_l\f^c+\frac{1}{12}(\f^6)^3\,F_{ij}^{fg}\,F_{kl}^{ab}\f^c
\right]\nonumber\\
&&\qquad\qquad\qquad
-\f^6\left[F_{ij}^{fg}D_k\f^aD_l\f^bD_m\f^c-\frac14\left(1-\frac43(\f^6)^2\right)
F_{ij}^{fg}\,F_{kl}^{ab}D_m\f^c\right]\bigg\}\,.\label{id15}
\eea
This expression is the $D=5$ analogue of \re{id1} in $D=4$, which likewise splits up in a gauge invariant and
a total divergence part.

Extracting the partial derivative $\pa_m$ from the gauge variant term in \re{vrv05}, one has
\be
\vep_{ijklm}\vep^{fgabc}\pa_m\f^6|\vec\f|^2\,\Xi_{ijkl}^{fgabc}=\vep_{ijklm}\vep^{fgabc}
\left\{\pa_m\left[\left(\f^6-\frac13(\f^6)^3\right)\Xi_{ijkl}^{fgabc}\right]-\left(\f^6-\frac13(\f^6)^3\right)
\pa_m\Xi_{ijkl}^{fgabc}\right\}\,,\label{id25}
\ee
which is the $D=5$ analogue of \re{id2} in $D=4$.

The question now is, does the divergence in the second term in \re{id25} result in a gauge invariant
quantity as was the case in \re{id2}, in the $D=4$?

The answer is found by calculating the divergence
\be
\label{divXi}
\vep_{ijklm}\vep^{fgabc}\,\pa_m\Xi_{ijkl}^{fgabc}=\frac14\vep_{ijklm}\vep^{fgabc}
\left[F_{ij}^{fg}F_{kl}^{ab}D_m\f^c-4\,(A_iA_j)^{fg}(A_kA_l)^{ab}A_m\f^c\right]\,,
\ee
which consists of a gauge invariant term, and the gauge variant term
\be
\vep_{ijklm}\vep^{fgabc}(A_iA_j)^{fg}(A_kA_l)^{ab}A_m\f^c\,,\label{gv}
\ee
which can be shown to be vanishing (see Appendix).

This results in the identity \re{id25} taking the desired form
\bea
\label{gvars}
\vep_{ijklm}\vep^{fgabc}\pa_m\f^6|\vec\f|^2\,\Xi_{ijkl}^{fgabc}&=&\vep_{ijklm}\vep^{fgabc}
\bigg\{\pa_m\left[\f^6\left(1-\frac13(\f^6)^2\right)\Xi_{ijkl}^{fgabc}\right]\nonumber\\
&&\qquad\qquad\qquad-\frac14\f^6\left(1-\frac13(\f^6)^2\right)
F_{ij}^{fg}|_{kl}^{ab}D_m\f^c\bigg\}\,,
\eea

Applying \re{id25} and \re{gvars} to \re{vrv05}, $(\vr_G-\vr_0)$ can be expressed 
in terms of $W$ and $\pa_i\Om_i$ as prescribed by \re{rGminusr0} with
\bea
W&=&\vep_{ijklm}\vep^{fgabc}\f^6\left[F_{ij}^{fg}D_k\f^aD_l\f^bD_m\f^c+\frac{5}{12}(\f^6)^2F_{ij}^{fg}F_{kl}^{ab}D_m\f^c\right]
\label{W5}\\
\Om_m&=&\f^6\pa_i\Om_{im}-\vep_{ijklm}\vep^{fgabc}\f^6\bigg[F_{ij}^{fg}\f^aD_k\f^bD_l\f^c+\frac{1}{12}(\f^6)^2
F_{ij}^{fg}F_{kl}^{ab}\f^c\nonumber\\
&&\qquad\qquad\qquad\qquad\qquad\qquad\qquad\qquad\qquad
+\left(1-\frac13(\f^6)^2\right)\Xi_{ijkl}^{fgabc}\label{Omm5}
\bigg]
\eea
which are the $D=5$ analogues of the $D=4$ expressions \re{W4} and \re{Oml4} for $W$ and $\Om_i$.

The result stated here coincides with that of Ref.~\cite{Callies:2013jbu} in the mathematical literature, which also covers dimensions
up to $D=5$.

\section{Summary and outlook}
The new result in this study is the implementation of the prescription \re{vr1}-\re{vr2} for the SCP in $D=5$.
To give as systematic as possible presentation, this has been presented in a unified manner together with all
so far known implementations, in $D=2,\ 3$ and $4$.

By far the most important application of SCP comes from the prescription \re{vr2}.
Since \re{vr2} is {\it essentially} total divergence 
(and {\it explicitly} total divergence in constraint-compliant parametrisation) it can be formally expressed as
\be
\vr^{(D)}=\pa_i(\om_i^{(D)}+\Om_i)\ ,\quad i=1,2,\dots,D\,.\label{totdiv}
\ee

Since \re{totdiv} is also gauge invariant, a Chern-Simons like density can be defined
in {\it one dimension lower}, in $d=D-1$ ``spacetime'',
\be
\label{SCS}
\Om_{\rm SCS}\stackrel{\rm def.}=\om_{i=D}^{(D)}+\Om_{i=D}
\ee
in which $\om_{i=D}^{(D)}$ here is the Wess-Zumino term. The density $\Om_{\rm SCS}$ in \re{SCS} referred to
as the Skyrme--Chern-Simons (SCS) density~\cite{Tchrakian:2015pka,Tchrakian:2021xzy} defined in $D-1$
``spacetime'' dimensions.
That the equations of motion of \re{SCS} are gauge invariant can be demonstrated 
exactly as in the case of the Chern-Simons density, since \re{vr2} (or \re{totdiv}) is gauge-invariant and
total-derivative like the Chern-Pontryagin (CP) density.

The idea of constructing a Chern-Simons like anomaly density, namely the Skyrme--Chern-Simons (SCS) density
{\it via} a one-step descent of the Skyrme--Chern-Pontryagin (SCP) density has appeared first in Ref.~\cite{Witten:1983tw}, for
descent of $5$ to $4$. Subsequently the special case with Abelian gauging was made concrete in Ref.~\cite{Callan:1983nx}.
There is a large body of mathematical literature applying equivariant cohomology to gauged Wess-Zumino models. See
Ref.~\cite{Figueroa-OFarrill:1994vwl} and references therein.

Both the anomaly density in \cite{Callan:1983nx} and the corresponding SCS density depend on the gauge field as well as the Skyrme
scalar, in contrast with the Chern-Simons density which is defined in terms of the gauge field only. Such an example for a SCS
density in $2+1$ dimensions is employed in \cite{Navarro-Lerida:2023fsr}, where it was shown that the effect of the SCS term
on the gauged Skyrmion is qualitatively the same as that of the CS term.

The quantitative study of the SCS anomaly in $3+1$ dimensions, both for Abelian and non-Abelian gauging, is under active consideration.

\section*{Acknowledgements}
The question whether the prescription \re{vr1}-\re{vr2} for the SCP density can be realised beyond $D=4$
was emphasised by V.~A.~Rubakov (R.I.P.). The present study is aimed at resolving this question. I am very
grateful to my colleague Eugen Radu for having read the manuscript carefully, and to him and to
Francisco Navarro-Lerida for extensive, valuable discussions on these topics. Thanks also to A~.P.~Balachandran,
Bjarke Gudnason, Derek Harland and Parameswaran Nair for useful discussions. Last but not least, my sincere thanks to
the referees of J. Phys. A, for their unstinting support and crucial advice.

\newpage
\begin{small}

\end{small}

\newpage 
\appendix
\setcounter{equation}{0}
\renewcommand{\theequation}{A.\arabic{equation}}
\section{Details of the calculation}
The $O(D+1)$ Skyrme scalar is $\f^A\ ,\ A=1,2,\dots,D+1$ is subject to
\be
|\f^A|^2=|\f^a|^2+(\f^{D+1})^2=1\ ,\quad a=1,2,\dots,D\,.\label{constrx}
\ee

The $SO(D)$ gauging prescription is
\bea
D_i\f^a&=&\pa_i\f^a+A_i\f^a\ ,\quad A_i\f^a=(A_i\f)^b=A_i^{ab}\f^b\label{coval22x}\\
D_i\f^{D+1}&=&\pa_i\f^{D+1}\label{covaA22x}.
\eea

$A_i^{ab}=-A_i^{ba}\ \ i=1,2,\dots,D\ ;a=1,2,\dots,D$ is the $SO(D)$ gauge connection, whose curvature is
\bea
F_{ij}^{ab}&=&\pa_{[i}A_{j]}^{ab}+(A_{[i}A_{j]})^{ab}\label{curv1x}\\
&\equiv&\pa_iA_j^{ab}+A_i^{ac}A_j^{cb}-(i,j)\nonumber\,.
\eea

For $D=5$ the winding number density is
\be
\label{wnd}
\vr_0=\vep_{ijklm}\,\vep^{abcdef}\,\pa_i\f^a\,\pa_j\f^b\,\pa_k\f^c\,\pa_l\f^d\,\pa_m\f^e\ \f^d
\ee
which is {\it total divergence} and {\it not gauge invariant}, and
replacing the partial derivatives $\pa_i\f^a$ in \re{wnd} with the covariant derivatives $D_i\f^a$
\be
\label{rG}
\vr_G=\vep_{ijklm}\,\vep^{abcdef}\,D_i\f^a\,D_j\f^b\,D_k\f^c\,D_l\f^d\,D_m\f^e\ \f^d
\ee
which is gauge invariant but {\it not total divergence} but is {\it gauge invarianct}~\footnote{
For all other $D=4,3,2$, the simpler versions of \re{wnd} and \re{rG} are obvious.}

The task is to calculate
\be
\label{dif}
\vr_G-\vr_0=\pa_i\Om_i[A,\f]-W[F,D\f]
\ee
where $W[F,D\f]$ is gauge invariant by construction and $\Om_i[A,\f]$ is gauge variant. (In increasing
dimensions, there is some arbitrariness in the splitting into these two terms.)

The relation \re{dif} is evaluated explicitly with
gauge group $SO(5)$, that defines the covariant derivative in \re{coval22x}. The
calculations are carried out directly, using the Leibniz rule and the tensor identities.

Collecting the gauge invariant pieces $\vr_G$ and $W$ in \re{dif}, and separately,
the individually gauge variant pieces $\vr_0$ and $\pa_i\Om_i$,
one has two equivalent definitions of a density
\bea
\vr&=&\vr_G+W[F,D\f]\label{vr1x}\\
&=&\vr_0+\pa_i\Om_i[A,\f]\,,\label{vr2x}
\eea
which is adopted as the definition for the density $\vr\stackrel{\rm def.}=\Om^{(d+1)}_{\rm SCP}$ presenting a lower
bound on the ``energy'' in the same way as does the usual CP density.

\bigskip

\noindent
{\bf\Large First step}

\bigskip

Using the tensor dentity
\be
\label{tensid}
V^b\,\vep^{{a_1}{a_2}a_3\dots {a_D}}=V^{a_1}\,\vep^{b\,a_2a_3\dots a_D}+V^{a_2}\,\vep^{a_1b\,a_3\dots a_D}
+V^{a_3}\,\vep^{a_1a_2\,b\dots a_D}+V^{a_D}\,\vep^{b\,a_2a_3\dots b}
\ee
in $D$ dimensnions, it can be seen that the quantities $(\vr_G-\vr_0)$ in dimensions $D=2,3,4,5$ reduce to
\bea
\frac12(\vr_G-\vr_0)&=&\frac12\,\vep_{ij}\,\vep^{fg}\,\pa_j\f^3\ A_i^{fg}
\equiv\vep_{ij}\,\pa_j\f^3\ A_i\label{rGr02y}\\
\frac13(\vr_G-\vr_0)&=&\vep_{ijk}\,\vep^{fga}\ \pa_k\f^4\ A_j^{fg}
\left(\pa_i\f^a+\frac12 A_i\f^a\right)\label{rGr03y}\\
\frac14(\vr_G-\vr_0)&=&\vep_{ijkl}\,\vep^{fgab}\ \pa_l\f^5\ A_k^{fg}
\left(\pa_i\f^a\pa_j\f^b+\pa_i\f^aA_j\f^b+\frac13\,A_i\f^aA_j\f^b\right)\label{rGr04y}\\
\frac15(\vr_G-\vr_0)&=&\vep_{ijklm}\,\vep^{fgabc}\ \pa_m\f^6\ A_l^{fg}
\bigg(2\pa_i\f^a\pa_j\f^b\pa_k\f^c+3\pa_i\f^a\pa_j\f^bA_k\f^c\nonumber\\
&&\qquad\qquad\qquad\qquad\qquad\qquad+2\pa_i\f^aA_j\f^bA_k\f^c+\frac12\,A_i\f^aA_j\f^bA_k\f^c\bigg)
\label{rGr05y}
\eea

The second step is to treat each case $D=2,3,4,5$ separately in a unified manner.

\bigskip

\noindent
{\bf\Large Second step}

\bigskip

It is convenient to express \re{rGr03y}-\re{rGr05y} formally as
\bea
-\frac1D(\vr_G-\vr_0)&=&\pa_n\f^{D+1}(\om_n^{(1)}+\om_n^{(2)})\label{sum}
\eea
in which $\om_n^{(1)}$ is encoded with terms like $A_i\f^a$ $exclusively$, and $\om_n^{(2)}$ by a $mixture$
of terms like $A_i\f^a$ $and$ $\pa_i\f^a$.

\subsection{$D=2$}
That \re{rGr02y} splits up in a total divergence term and a gauge invariant term is seen immediately.
Applying the Leibniz rule once, \re{rGr02y} can be immediately be expressed as
\be
\label{rGr02ys}
\frac12(\vr_G-\vr_0)=\vep_{ij}\left[\pa_j(\f^3\ A_i)-\frac12\,\f^3\,F_{ij}\right]\,,
\ee
consisting of a total divergence  and a gauge invariant term.

\subsection{$D=3$}
In this case
\bea
\om_k^{(1)}&=&\frac12\,\vep_{ijk}\vep^{fga}\,A_j^{fg}\,A_i\f^a\nonumber\\
&=&\frac12\,\vep_{ijk}\vep^{fga}(A_{[i}A_{j]})^{fg}\,\f^a
\label{om31}\\
\om_k^{(2)}&=&\vep_{ijk}\vep^{fga}\,A_j^{fg}\,\pa_i\f^a\nonumber\\
&=&\pa_i\Om_{ik}-\frac12\,\vep_{ijk}\vep^{fga}\,\f^a\,\pa_{[i}A_{j]}^{fg}
\label{om32}
\eea
where, \re{om31} has resulted from using the identity \re{tensid} for $D=3$ and  $\Om_{ik}$ in \re{om32} 
is defined as
\be
\label{Omiky}
\Om_{ik}=\vep_{ijk}\vep^{fga}\,(A_j^{fg}\,\f^a)\,.
\ee

Thus adding $\om_k^{(1)}$ and $\om_k^{(2)}$ as in \re{sum}, which are now given by \re{om31} and \re{om32},
results in
\be
\label{result3y}
\frac13(\vr_G-\vr_0)=\pa_k\f^4\left[\pa_i\Om_{ik}-\frac12\,\vep_{ijk}\vep^{fga}\,\f^a\,F_{ij}^{fg}\right]
\ee
which consists of a part that is gauge invariant, and a total divergence term which is gauge variant, as seen in
\be
\label{result3x}
\frac13(\vr_G-\vr_0)=\pa_k\left[\f^4\pa_i\Om_{ik}\right]-\frac12\,\pa_k\f^4\,\vep_{ijk}\vep^{fga}\,\f^a\,F_{ij}^{fg}
\ee
or alternatively,
\bea
\frac13(\vr_G-\vr_0)&=&
\pa_k\left[\f^4\pa_i\Om_{ik}-\frac12\,\f^4\,\vep_{ijk}\vep^{fga}\,\f^a\,F_{ij}^{fg}\right]
+\frac12\,\vep_{ijk}\vep^{fga}\f^4\,D_k\f^a\,F_{ij}^{fg}\label{result3z}\\
&\equiv&\pa_k\Om_k+\frac12\,\vep_{ijk}\vep^{fga}\f^4\,D_k\f^a\,F_{ij}^{fg}\,,\label{result3zs}
\eea
in the second line of which the vector valued density $\Om_k$ is preciselythe $D=3$ case of $\Om_i$ appearing in \re{vr2} .

\subsection{$D=4$}
According to the notation in \re{sum}, $\om_l^{(1)}$ and $\om_l^{(2)}$ can be read from \re{rGr04y} as,
\bea
\om_l^{(1)}&=&\frac13\,\vep_{ijkl}\vep^{fgab}\,A_k^{fg}\,A_i\f^aA_j\f^b\nonumber\\
&=&\frac13\,\vep_{ijkl}\vep^{fgab}(A_iA_j)^{fg}\left(3\f^aA_k\f^b+\frac12|\vec\f|^2A_k^{ab}\right)\,,\qquad
|\vec\f|^2=|\f^a|^2\label{om41}
\eea
\bea
\om_l^{(2)}&=&\vep_{ijkl}\vep^{fgab}\,A_k^{fg}\,\pa_i\f^a\,D_j\f^b\nonumber\\
&=&\pa_i\Om_{il}^{(1)}+\vep_{ijkl}\vep^{fgab}\,\pa_iA_j^{fg}\,\f^aD_j\f^b-\om_l^{(a)}-\om_l^{(b)}\label{42}
\eea
 where
\bea
\Om_{il}^{(1)}&=&\vep_{ijkl}\vep^{fgab}\,A_k^{fg}\,\f^aD_j\f^b\label{Om14}\,.
\eea

The quantities $\om_l^{(a)}$ and $\om_l^{(b)}$ in \re{42} are calculated to give
\bea
\om_l^{(a)}&=&\vep_{ijkl}\vep^{fgab}\,A_k^{fg}\,\f^a\,\pa_iA_j\f^b\nonumber\\
&=&-\vep_{ijkl}\vep^{fgab}\,\pa_iA_j^{fg}\left(\f^aA_k\f^b+\frac12|\vec\f|^2\,A_k^{ab}\right)\label{oma4}
\eea
and
\bea
\om_l^{(b)}&=&\vep_{ijkl}\vep^{fgab}\,A_k^{fg}\,\f^a\,A_j\pa_i\f^b\nonumber\\
&=&\frac12\pa_i\Om_{il}^{(2)}+\vep_{ijkl}\vep^{fgab}\left[\pa_iA_j^{fg}\left(\f^aA_k\f^b
+\frac14|\vec\f|^2\,A_k^{ab}\right)-(A_iA_j)^{fg}\f^a\pa_k\f^b\right]\,,\label{omb4}
\eea
the density $\Om_{il}^{(2)}$ being
\bea
\Om_{il}^{(2)}&=&\vep_{ijkl}\vep^{fgab}\,A_k^{fg}\,\f^aA_j\f^b\label{Om24}\,.
\eea

The analogue of the density $\Om_{ik}$ \re{Omiky} for $D=3$, can be denoted as
\bea
\Om_{il}&=&(\Om_{il}^{(1)}-\frac12\Om_{il}^{(2)})=\vep_{ijkl}\vep^{fgab}A_k^{fg}\,\f^a
\left(\pa_j\f^b-\frac12A_j\f^b\right)\,.\label{tildeOm4y}
\eea

Substituting \re{oma4} and \re{omb4} in \re{42}, $\om_l^{(2)}$ is
\bea
\om_l^{(2)}&=&\pa_i\Om_{il}+\vep_{ijkl}\vep^{fgab}\left\{\pa_iA_j^{fg}\f^aD_k\f^b
+(A_iA_j)^{fg}\f^a\pa_k\f^b+\frac14|\vec\f|^2\pa_iA_j^{fg}\,A_k^{ab}\right\}\label{om42}\,.
\eea

Adding \re{om41} and \re{om42} yields $(\om_l^{(1)}+\om_l^{(2)})$, and contracting this with
$\pa_l\f^5$, we have $(\vr_G-\vr_0)$ as $per$ \re{sum}
\be
\frac14(\vr_G-\vr_0)=\pa_l{\f^5\pa_i\Om_{il}}+\frac12\,\vep_{ijkl}\vep^{fgab}\pa_l\f^5\left\{
F_{ij}^{fg}\,\f^aD_k\f^b+\frac12|\vec\f|^2\Xi_{ijk}^{fgab}(4)\right\}
\label{vrv04y}
\ee
where
\be
\label{Xi4z}
\Xi_{ijk}^{fgab}(4)=A_k^{ab}\left[\pa_iA_j^{fg}+\frac23(A_iA_j)^{fg}\right]\,,
\ee
and the $l=4$ component of
\be
\Xi_l^{(4)}=\vep_{ijkl}\vep^{fgab}\Xi_{ijk}^{fgab}(4)\ ,\label{Xil4}
\ee
can be identfied with the Chern-Simons density in three dimensions.

The expression \re{vrv04y} for $(\vr_G-\vr_0)$ is at the same stage as that of its $D=3$
counterpart \re{result3y}, but unlike the former, it is not yet 
a sum of a total-divergence and gauge invariant term.
To render it to that form one needs to extract the partial derivative $\pa_l$ in \re{vrv04y}.

Extracting the partial derivative from the gauge invariant term in \re{vrv04y}
is optional but useful. It yields
\bea
\vep_{ijkl}\vep^{fgab}\pa_l\f^5\,\{F_{ij}^{fg}\,\f^aD_k\f^b\}&=&\vep_{ijkl}\vep^{fgab}\bigg\{\pa_l
\left[\f^5\,F_{ij}^{fg}\,\f^aD_k\f^b\right]\nonumber\\
&&\qquad\qquad-\f^5\left[F_{ij}^{fg}D_k\f^aD_l\f^b+\frac18|\vec\f|^2F_{ij}^{fg}F_{kl}^{ab}\right]\bigg\}\,.\label{id1y}
\eea

Carrying out the same for the gauge-variant term yields
\bea
\vep_{ijkl}\vep^{fgab}|\vec\f|^2\pa_l\f^5\Xi_{ijk}^{fgab}(4)
&=&\vep_{ijkl}\vep^{fgab}\bigg\{\pa_l\left[(\f^5-\frac13(\f^5)^3)\Xi_{ijk}^{fgab}(4)\right]\nonumber\\
&&\qquad\qquad\quad +\frac14(\f^5-\frac13(\f^5)^3)F_{ij}^{ab}F_{kl}^{cd}\bigg\}\label{id2y}
\eea
one turns up with the final result in the required form
\be
\frac14\,(\vr_G-\vr_0)=\pa_l\Om_l+\frac12\,\vep_{ijkl}\vep^{fgab}\,\f^5\left\{F_{ij}^{fg}D_k\f^aD_l\f^b+
\frac{1}{12}(\f^5)^2\,F_{ij}^{fg}F_{kl}^{ab}\right\}\label{rGr04x}
\ee
where the vector valued density $\Om_l$ is given by
\be 
\label{Oml4y}
\Om_l=\f^5\pa_i\Om_{il}+\vep_{ijkl}\vep^{fgab}\left\{\frac12\f^5\,F_{ij}^{fg}\f^aD_k\f^b
+\frac14\left(\f^5-\frac13(\f^5)^3\right)\Xi_{ijk}^{fgab}(4)\right\}
\ee
which is precisely the $D=4$ case of $\Om_i$ appearing in \re{vr2} .
\subsection{D=5}
According to the notation in \re{sum}, $\om_l^{(1)}$ and $\om_l^{(2)}$ can be read from \re{rGr05y} as,
\bea
\om_m^{(1)}&=&\frac12\,\vep_{ijklm}\vep^{fgabc}\,A_l^{fg}\,A_i\f^aA_j\f^bA_k\f^c\nonumber\\
&=&-2\,\vep_{ijklm}\vep^{fgabc}(A_iA_j)^{fg}\,A_l\f^c\left(\f^aA_k\f^b+\frac14|\vec\f|^2\,A_k^{ab}\right)
\label{om15}
\eea
and
\bea
\om_m^{(2)}&=&\pa_i\Om_{im}^{(1)}-\vep_{ijklm}\vep^{fgabc}
\pa_iA_j^{fg}\f^a(2D_k\f^bD_l\f^c-\pa_l\f^cA_k\f^b)-\om_m^{(a)}-3\,\om_m^{(b)}-4\,\om_m^{(c)}\label{om25}
\eea
where
\bea
\Om_{im}^{(1)}&=&\vep_{ijklm}\vep^{fgabc}A_l^{fg}\f^a(2D_j\f^bD_k\f^c-\pa_j\f^bA_k\f^c)\label{Om15}
\eea
and
\bea
\om_m^{(a)}&=&\vep_{ijklm}\vep^{fgabc}A_k^{fg}\f^a(3\pa_l\f^b+4A_l\f^b)\pa_iA_j\f^c\label{oma5}\\
\om_m^{(b)}&=&\vep_{ijklm}\vep^{fgabc}A_k^{fg}\f^a\pa_l\f^b\,A_j\pa_i\f^c\label{omb5}\\
\om_m^{(c)}&=&\vep_{ijklm}\vep^{fgabc}A_k^{fg}\f^aA_l\f^b\,A_j\pa_i\f^c\label{omc5}
\eea

Here, $\om_m^{(a)}$, $\om_m^{(b)}$ and $\om_m^{(c)}$ must be calculated individually, as was done above for
in the $D=4$ case above for $\om_l^{(a)}$ and $\om_l^{(b)}$ in \re{oma4} and \re{omb4} respectively,.

The results of these calclations are listed here:
\be
\label{oma5x}
\om_m^{(a)}=\vep_{ijklm}\vep^{fgabc}\,\pa_iA_j^{fg}\left(\f^aA_k\f^b+\frac12|\vec\f|^2\,A_k^{ab}\right)
(3\,\pa_l\f^c+4\,A_l\f^c)
\ee
\bea
\om_m^{(b)}&=&\frac13\,\pa_i\Om_{im}^{(2)}+\frac23\,\vep_{ijklm}\vep^{fgabc}\bigg\{
-\pa_iA_j^{fg}\,\pa_l\f^c\left(\f^aA_k\f^b+\frac14|\vec\f|^2\,A_k^{ab}\right)\nonumber\\
&&\qquad\qquad\qquad\qquad\qquad\qquad+(A_iA_j)^{fg}\,\f^a\pa_k\f^b\pa_l\f^c\bigg\}\label{omb5x}
\eea
where
\be
\label{Om25}
\Om_{im}^{(2)}=\vep_{ijklm}\vep^{fgabc}A_l^{fg}\f^a\,\pa_j\f^bA_k\f^c
\ee
and
\bea
\om_m^{(c)}&=&\frac13\pa_i\Om_{im}^{(3)}+\frac13\,\vep_{ijklm}\vep^{fgabc}\bigg\{
-\pa_iA_j^{fg}\,A_l\f^c\left(3\f^aA_k\f^b+|\vec\f|^2\,A_k^{ab}\right)\nonumber\\
&&\qquad\qquad\qquad\qquad\qquad\qquad+(A_iA_j)^{fg}\,\pa_l\f^c\left(3\f^aA_k\f^b+\frac12|\vec\f|^2\,A_k^{ab}\right)
\label{omc5x}
\eea
where
\be
\label{Om35}
\Om_{im}^{(3)}=\vep_{ijklm}\vep^{fgabc}A_l^{fg}\f^a\,A_j\f^bA_k\f^c
\ee

Substituting \re{oma5x}, \re{omb5x} and \re{omc5x} into \re{om25}

\bea
\om^{(2)}_m&=&\pa_i\Om_{im}-2\vep_{ijklm}\vep^{fgabc}\bigg\{\pa_iA_j^{fg}
\left[\f^a\,D_k\f^bD_l\f^c+\frac12\,|\vec\f|^2\,A_k^{ab}\left(\pa_l\f^c+\frac23\,A_l\f^c\right)\right]\nonumber\\
&&\qquad\qquad\qquad\qquad\quad+(A_iA_j)^{fg}\left[\f^a\pa_k\f^b(\pa_l\f^c+2\,A_l\f^c)+\frac13\,
|\vec\f|^2\,A_k^{ab}\,\pa_l\f^c\right]\bigg\}\label{om25x}
\eea
where
\bea
\label{tildeOm5y}
\Om_{im}&=&\Om_{im}^{(1)}-\Om_{im}^{(2)}-\frac43\Om_{im}^{(3)}\nonumber\\
&=&2\vep_{ijklm}\vep^{fgabc}\,A_l^{fg}\,
\f^a(\pa_j\f^b\pa_k\f^c+\pa_j\f^bA_k\f^c+\frac13\,A_j\f^bA_k\f^c)
\eea
in which $\Om_{im}^{(1)}$, $\Om_{im}^{(2)}$ and $\Om_{im}^{(3)}$ are given in \re{Om15}, \re{Om25} and \re{Om35}
respectively, and $\Om_{im}$ is the $D=5$ analogue of $\Om_{il}$ given by \re{tildeOm4y} in $D=4$.

Adding $\om_m^{(1)}$ given in \re{om15}, to $\om_m^{(2)}$ given in \re{om25x},
\bea
\om_m^{(1)}+\om_m^{(2)}&=&\pa_i\Om_{im}-\vep_{ijklm}\vep^{fgabc}\left\{
F_{ij}^{fg}\,\f^a\,D_k\f^bD_l\f^c+|\vec\f|^2\,\Xi_{ijkl}^{fgabc}(5)\right\}\label{om1plusom2}
\eea
where
\bea
\Xi_{ijkl}^{fgabc}(5)&=&A_k^{ab}\left[(\pa_l\f^c+\frac23\,A_l\f^c)\,\pa_iA_j^{fg}
+\frac23(\pa_l\f^c+\frac34\,A_l\f^c)(A_iA_j)^{fg}\right]\,,\label{CS5y}
\eea
which is the $5$-dimensional analogue of \re{Xi4z} in $4$ dimensions, such that the $m=5$ component of
\be
\Xi_m^{(5)}=\vep_{ijklm}\vep^{fgabc}\,\Xi_{ijkl}^{fgabc}(5)\label{Xim5}
\ee
is the five dimensional analogue of the term  $\Xi_l^{(4)}$ in four dimensions, appearing in \re{Xil4}.

As $per$ \re{sum}, \re{om1plusom2} implies
\bea
\frac15(\vr_G-\vr_0)&=&\pa_m\f^6\pa_i\Om_{im}-\vep_{ijklm}\vep^{fgabc}\pa_m\f^6\left\{
F_{ij}^{fg}\,\f^a\,D_k\f^bD_l\f^c+|\vec\f|^2\,\Xi_{ijkl}^{fgabc}(5)\right\}\label{vrv05y}
\eea

This expression is the $D=5$ analogue of \re{vrv04y} in $D=4$.
The first term, $\pa_m(\f^6\pa_i\Om_{im})$, is a total divergence . Inside the chain
bracket the first term is gauge invariant, but the second term in front of $|\vec\f|^2$ is neither 
gauge invariant nor total divergence, which is unsatisfactory from the viewpoint of casting this expression 
in the form \re{dif}, in terms of a gauge invariant quantity $W$ and a total divergence $\pa_i\Om_i$.
This can be effected by pulling out~\footnote{Strictly speaking this does not have to involve the first
term inside the chain bracket of \re{vrv05y} since that term is gauge invariant as it stands. But we opt to
do that since that results in a more useful expression for later application of Bogomol'nyi type inequalities.}
the partial derivative $\pa_m$
as was done in the $D=4$ case.

But this case being more complex than the $D=4$, it is helpful to treat the gauge
invariant term and the gauge variant term (in front of $|\vec\f|^2$) separately.

Pulling out $\pa_m$ from the gauge invariant term in \re{vrv05y} results in the expression
\bea
&&\vep_{ijklm}\vep^{fgabc}\pa_m\f^6\,F_{ij}^{fg}\,\f^a\,D_k\f^bD_l\f^c=\vep_{ijklm}\vep^{fgabc}\cdot\nonumber\\
&&\qquad\qquad
\cdot\bigg\{\pa_m\left[\f^6\,F_{ij}^{fg}\f^aD_k\f^bD_l\f^c+\frac{1}{12}(\f^6)^3\,F_{ij}^{fg}\,F_{kl}^{ab}\f^c
\right]\nonumber\\
&&\qquad\qquad\qquad
-\f^6\left[F_{ij}^{fg}D_k\f^aD_l\f^bD_m\f^c+\frac14\left(1-\frac23(\f^6)^2\right)
F_{ij}^{fg}\,F_{kl}^{ab}D_m\f^c\right]\bigg\}\,.\label{ginv}
\eea
This expression is the $D=5$ analogue of \re{id1y} in $D=4$.

Doing the same with the gauge variant term results in
\be
\label{gvar}
\vep_{ijklm}\vep^{fgabc}\pa_m\f^6|\vec\f|^2\,\Xi_{ijkl}^{fgabc}(5)=\vep_{ijklm}\vep^{fgabc}
\left\{\pa_m\left[\left(\f^6-\frac13(\f^6)^3\right)\Xi_{ijkl}^{fgabc}\right]-\left(\f^6-\frac13(\f^6)^3\right)
\pa_m\Xi_{ijkl}^{fgabc}(5)\right\}\,,
\ee
in which the divergence term in the RHS is calculated to be
\be
\label{divXiy}
\vep_{ijklm}\vep^{fgabc}\,\pa_m\Xi_{ijkl}^{fgabc}(5)=\frac14\vep_{ijklm}\vep^{fgabc}
\left[F_{ij}^{fg}F_{kl}^{ab}D_m\f^c-4\,(A_iA_j)^{fg}(A_kA_l)^{ab}A_m\f^c\right]\,,
\ee
which is the counterpart of the divergence
\be
\label{divCS}
\vep_{ijkl}\vep^{fgab}\,\pa_l\Xi_{ijk}^{fgab}(4)=
-\frac14\vep_{ijkl}\vep^{fgab}
\,F_{ij}^{fg}F_{kl}^{ab}\,,
\ee
occuring in the $D=4$ case, which, as it stands is gauge invariant.

Likewise here, the divergence \re{divXiy} is also gauge invariant since the gauge-variant term in it can be cast in the form
\be
\label{gvy}
\vep_{ijklm}\vep^{fgabc}\,(A_iA_j)^{fg}(A_kA_l)^{ab}A_m\f^c=-4\,\vep_{ijklm}\vep^{fgabc}\,(A_iA_jA_k)^{fg}\,(A_lA_m)^{ab}\,\f^c\,,
\ee
which vanishes identically, by symmetry.

It is seen that the calcululus presented here for the explicit construction of $SO(D)$ gauged SCP densities
proceeds with a uniform pattern in dimensions $D=2,3,4,5$, implying its extension to arbitrary dimension $D$.

\end{document}